\documentclass[12pt]{article}
\pdfoutput=1
\usepackage{lscape}
\usepackage{multirow}
\usepackage{color}
\usepackage{bm,bbm}
\usepackage{marvosym}
\usepackage{amssymb,amsmath}
\usepackage{amstext}
\usepackage{amscd}
\usepackage{graphicx}
\usepackage{shadow}
\usepackage[all]{xy}
\usepackage{hyperref}
\usepackage{yfonts}
\usepackage{latexsym,feynmf}

\newcommand{\eqn}[1]{(\ref{#1})}

\newcommand{\braket}[2]{ \langle #1 \lvert #2 \rangle }  
\newcommand{\ketbra}[2]{ \lvert #1 \rangle \,\langle #2 \rvert }  

\def\be{\begin{equation}}
\def\ee{\end{equation}}
\def\beq{\begin{equation}}
\def\eeq{\end{equation}}
\def\bea{\begin{eqnarray}}
\def\eea{\end{eqnarray}}

\def\nn{\nonumber}

\def\sideremark#1{\ifvmode\leavevmode\fi\vadjust{\vbox to0pt{\vss
 \hbox to 0pt{\hskip\hsize\hskip1em
 \vbox{\hsize3cm\tiny\raggedright\pretolerance10000
  \noindent #1\hfill}\hss}\vbox to8pt{\vfil}\vss}}}

\DeclareMathAlphabet{\mathpzc}{OT1}{pzc}{m}{it}

\begin{document}

\thispagestyle{empty}

\begin{fmffile}{QQQFeynmf}{
\fmfset{zigzag_width}{1thick}
\fmfset{dot_len}{0.75mm}
\fmfcmd{%
  style_def majorana expr p =
    cdraw p;
    cfill (harrow (reverse p, .25));
    cfill (harrow (p, .25))
  enddef;}
\fmfcmd{vardef endpoint_dot expr p =
  save oldpen; pen oldpen;
  oldpen := currentpen;
  pickup oldpen scaled 4;
  cdrawdot p;
  pickup oldpen;
enddef;}
\fmfcmd{style_def deld expr p =
  cdraw p;
  endpoint_dot point 0.25 of reverse p;
 enddef;}
\fmfcmd{style_def ddeld expr p =
  cdraw p;
  endpoint_dot point 0.25 of reverse p;
  endpoint_dot point 0.25 of p;
 enddef;}
\fmfcmd{style_def ddel expr p =
  cdraw p;
  endpoint_dot point 0.25 of p;
 enddef;}

\vspace{.8cm}
\setcounter{footnote}{0}
\begin{center}
\vspace{-50mm}
{\Large
 {\bf Is Quantum Gravity a Chern--Simons Theory? }\\[4mm]

 {\sc \small
    R.~Bonezzi$^{\mathfrak B}$,  O.~Corradini$^{\mathfrak C}$  and A.~Waldron$^{\mathfrak W}$}\\[3mm]

{\em\small
                      ~${}^\mathfrak{B}$
Dipartimento di Fisica,\\ Universit\`a di Bologna, via Irnerio 46, I-40126 Bologna, Italy   \\
and INFN sezione di Bologna, via Irnerio 46, I-40126 Bologna, Italy\\
{\tt bonezzi@bo.infn.it}\\[1mm]
          ~${}^\mathfrak{C}$
          Facultas de Ciencias en F\'isica y Matem\'aticas,\\ Universidad Aut\'onoma de Chiapas,
          Ciudad Universitaria, Tuxtla Guti\'errez 29050, M\'exico\\
          and Dipartimento di Scienze Fisiche, Informatiche e Matematiche,\\ Universit\`a di Modena e Reggio Emilia,
          Via Campi 213/A,  I-41125 Modena, Italy\\ 
 {\tt olindo.corradini@unach.mx}\\[1mm]
~${}^{\mathfrak W}\!$
            Department of Mathematics\
            University of California,
            Davis CA 95616, USA\\
            {\tt wally@math.ucdavis.edu}

            }
 }

\bigskip

{\sc Abstract}\\[1mm]

\end{center}

\noindent
We propose a model of quantum gravity in arbitrary dimensions defined in terms of
the BV~quantization of a supersymmetric, infinite dimensional  matrix model. This gives an (AKSZ-type) Chern--Simons theory with gauge algebra the space of observables of a quantum mechanical Hilbert space~$\cal H$.
The model is motivated by previous attempts to formulate gravity in terms of non-commutative, phase space, field theories  as well as the Fefferman--Graham curved analog of Dirac spaces for conformally invariant wave equations.
The field equations are
flat connection conditions amounting to  zero curvature and parallel conditions on operators acting on~${\cal H}$.
This matrix-type model may give a better defined setting for a quantum gravity path integral.
We demonstrate that its underlying physics is a summation over Hamiltonians labeled by a conformal class of metrics and thus a sum over causal structures. This gives in turn a model
summing over fluctuating metrics plus a tower of additional modes---we speculate that these could yield improved UV behavior.





\newpage



\section{Introduction}\label{intro}

The problem of unifying quantum mechanics and gravity has vexed physicists  since  the early twentieth century. However, the absence of hard experimental data at scales where quantum gravity effects are expected to dominate has meant that even knowing the physical questions a unified model should answer has been difficult.
On the other hand, given the spectacular success of classical general relativity which was discovered on the basis of Einstein's brilliant theoretical and mathematical insight, hope that its quantization could be understood by theoretical methods has never been abandoned. Indeed, the major original stumbling block---non-renormalizability of gravity treated as the quantum field theory of a massless spin~2 particle---is solved  by the leading quantum gravity candidate--String Theory. Moreover, by
aiming for a grand unification of particle physics, gravity and quantum mechanics, String Theory in principle applies to physical settings probed by collider experiments. Although String Theory even has standard model-like solutions, it is currently believed to suffer from a massive loss of predictivity due to a vast landscape of vacua that, for the moment at least,  has forced anthropic reasoning to the fore. It is therefore interesting to investigate other models, that like String Theory, predict the presence of gravity. We present one such model in this Article.

The aim of physics is to predict the outcome of experiments based on a minimal set of fundamental laws. A basic physical construct is therefore a set of spacetime events which are typically modeled by a spacetime manifold.
Often this spacetime is equipped with a \mbox{(pseudo-)Riem\-annian} metric. Our first premise is that a causal structure (or in geometrical terms a conformal class of metrics) is more fundamental than a Riem\-annian metric. In its most basic formulation our model is not written in these terms, but we will demonstrate that it does predict a sum over causal structures. Rather, as basic input, we demand only a choice of quantum mechanical Hilbert space. This should be thought of analogously to the single particle Hilbert space of a quantum field theory. In standard quantum mechanics, a Hamiltonian governing dynamics is also a required input, our proposal however is that the {\it r\^ole}
of a quantum gravity theory is to give dynamics to the space of all possible quantum mechanical Hamiltonians.

Let us now give the ancestral history of our model, which we will
define in the next Section. Its genesis is Dirac's discovery that  conformally invariant wave equations in four dimensional Minkowski space could be reformulated in a six dimensional spacetime with two timelike directions~\cite{Dirac:1936fq}. This is in fact the Lorentzian version of what is known as the flat model for a conformal geometry,\footnote{Note that the term conformal invariance is employed in the physics literature to indicate invariance under the conformal isometries of a background spacetime, while in the mathematics literature it refers to symmetry under local rescalings of the metric (Weyl invariance in physics parlance).} see Figure~\ref{rays}.

\begin{figure}
\begin{center}
\includegraphics[height=6.5cm,width=7cm]{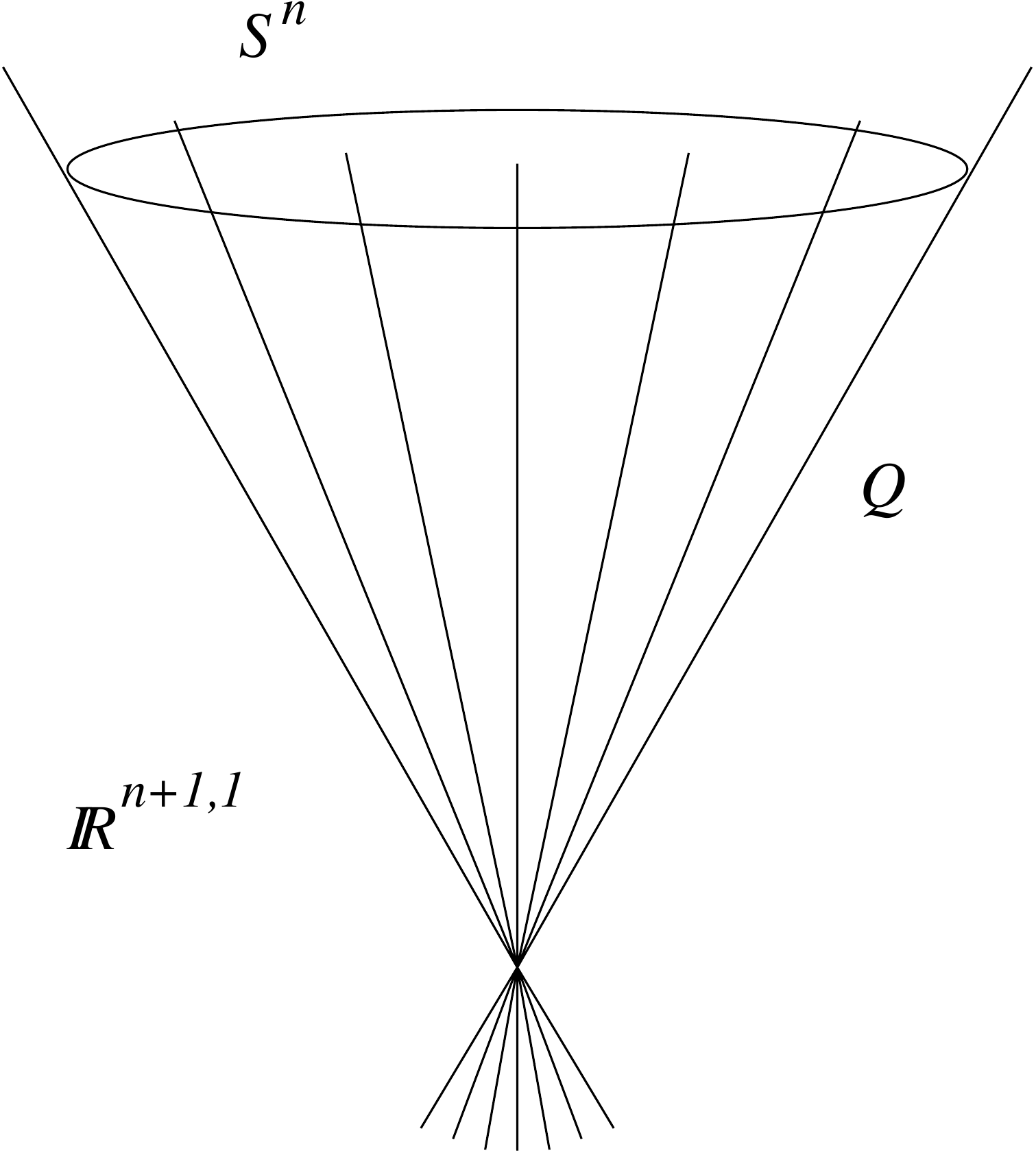}
\end{center}
\caption{The flat model for a conformal manifold. The~$n$-sphere with its canonical  conformal class of metrics is given by the space of lightlike rays~$Q$ in~${\mathbb R}^{n+1,1}$.\label{rays}}
\end{figure}

The next major ingredient is a curved analogue of the Dirac space. In a seminal paper, Fefferman and Graham showed that
ambient~$(d+2)$-dimensional metrics~$g_{MN}$ obeying a closed homothety condition\footnote{Observe that this equation is the real analog of the condition that K\"ahler metrics derive from a K\"ahler potential, since it implies~$g_{MN}=\frac12 \nabla_M \nabla_N X^2$ where~$X^2$ is the defining function for the curved analogue of the Dirac cone.}
\begin{equation}\label{homothety}
g_{MN}=\nabla_M X_N
\end{equation}
describe~$d$-dimensional conformal geometries on an underlying conformal manifold~$(M,[g])$~\cite{FG}. (The~$(d+2)$-dimensional geometry~$(\tilde M,g_{MN})$ is called a Fefferman--Graham (FG) ambient space and has signature~$(p+1,q+1)$ for signature~$(p,q)$ conformal geometries.) In fact, Fefferman and Graham also constructed asymptotic expansions of Ricci flat solutions for~$g_{MN}$; these underly the FG expansions for asymptotically AdS metrics relied upon by the AdS/CFT correspondence (we will not require, by definition, that FG metrics obey a Ricci flat condition).

The problem of finding conformal invariants and conformally invariant operators is more difficult than the analogous one for diffeomorphisms. Important progress was made by Graham, Jennes, Mason and Sparling (GJMS) who realized that the FG ambient space admitted an~$\mathfrak{sl}(2)$ algebra of differential operators
\begin{equation}\label{GJMS}
\left\{X^2\, ,\ \nabla_X +\frac{d+2}{2}\, ,\ \Delta\right\}\, ,
\end{equation}
and that these could be used to generate conformally invariant operators  whose leading symbol is given by powers of the~$d$-dimensional Laplacian~\cite{GJMS}. The space of all such triples of operators, which we dub a {\it GJMS algebra}, will play a crucial {\it r\^ole} in the following. To study its physics applications we  need to understand why conformal geometries grant Einstein manifolds\footnote{Recall that an Einstein manifold is one whose Einstein tensor is proportional to the metric (in other words these are solutions of cosmological Einstein gravity).}  a distinguished mantle.

Tensors on the FG ambient space, classified by weight (the eigenvalue of~$\nabla_X$), and defined up to equivalence along the cone\footnote{We use the notation~${\cal Z}$ for the zero locus of a function.}~$Q:={\cal Z}(X^2)$, {\it i.e.} \begin{equation}\label{coneE} T\sim T + X^2 S\, ,\end{equation} for smooth tensors~$T$ and~$S$, are known as tractors. These are equivalent to sections of the so-called tractor bundle along the underlying conformal manifold~$M$. These vector bundles were first formalized by Bailey, Eastwood and Gover~\cite{BEG} in order to generalize Penrose's twistor construction~\cite{Penrose} to arbitrary dimensions. The tractor bundle~${\cal T}M$ comes equipped with a canonical (tractor) connection which is crucial for an extremely important result:~${\cal T}M$  admits a parallel section,
\begin{equation}\label{parallel}
\nabla I^M = 0\, ,
\end{equation}
iff the conformal manifold is conformally Einstein~\cite{BEG}. This result is constructive; it determines the Weyl rescaling required to bring a given metric in the conformally Einstein class of metrics to an Einstein one in terms of the parallel {\it scale tractor}~$I^M$. In fact, the scale tractor provides the link between physics and conformal geometry: The dynamics of---not necessarily conformally invariant---physical systems is given by evolution along the ambient vector field~$I^M$.

This sets the stage for a crucial observation, first made by Marnelius and then extended to a new physics rubric  by Bars and collaborators: The ambient space of a Lorentzian space time has signature~$(-,-,+,+,\cdots)$ and thus two timelike directions. Thus, in what was dubbed~$2T$-physics, they studied the analog of a relativistic particle moving in a spacetime with two timelike directions, subject to not one mass-shell constraint, but an~$\frak{sl}(2)$ triplet of first class constraints~\cite{Marnelius:1978fs,Bars}. In this context, it is enlightening to view~$ \frak{sl}(2)$  as either~$\frak{so}(2,1)=\frak{co}(\mathbb R)$ or~${\frak {sp}}(2)$. From the former viewpoint, one is gauging the worldline conformal group, while latter manifests a Howe dual pair~\cite{Howe} of the ambient symplectic group
\begin{equation}\label{dual}
\frak{sp}(2)\otimes \frak{so}(d,2) \subset \frak{sp}\big(2(d+2)\big)\, .
\end{equation}
In the above, the two algebras on the left hand side are maximal cocommutants so that gauging~$\frak{sp}(2)$ in~$(d+2)$-dimensional quantum mechanics guarantees a remaining (but possibly hidden) conformal symmetry~$\frak{so}(d,2)$.
Different worldline gauge choices give various~$d$-dimensional theories (or ``shadows'', {\it e.g.} the relativistic particle~\cite{Marnelius:1978fs}, the hydrogen atom and harmonic oscillator, to name a few of these surprisingly dual theories~\cite{BarsShadows}) from the same~$(d+2)$-dimensional models and thus establish a string of dualities between models with~$\frak{so}(d,2)$-symmetry~\cite{Araya:2013bca}.
A key problem, therefore, was to second quantize this model, the original hope being that this might give a unified model for duality symmetries. The crucial observation of~\cite{BarsRey} was that this could be achieved via a non-commutative field theory with fields living on the phase space of the ~$(d+2)$-dimensional ambient manifold governed by a Chern--Simons action.
A remarkable feature of this model is that it can be used to describe gravity.

The dynamics of the model in~\cite{{BarsRey}} amounts to finding all triplets of quantum mechanical Hamiltonians obeying an~$\frak{sl}(2)$ algebra.  The classical version of this problem was solved in~\cite{BarsSol} and subsequently quantized in~\cite{Bonezzi}. These Hamiltonians are described by {\it conformal geometry moduli} consisting of an FG ambient metric as in~\eqn{homothety} and a (tractor) Maxwell gauge field~\cite{Bars:2008sz}. A proposal how to obtain gravity
from this data was given in~\cite{BarsBRST} based on BRST reasoning and results for ``$2T$-gravity'' actions. This amounts to imposing the three Schr\"odinger equations (more strictly Hamiltonian constraints) for  each of  these Hamiltonians and integrating over the conformal geometry moduli. This computation was performed in~\cite{Bonezzi} using  tractor calculus~\cite{BEG} and in particular the parallel condition~\eqn{parallel} (an earlier $2T$ gravity approach was proposed and studied in~\cite{Bars:2008sz} which amounts essentially to rewriting the Einstein--Hilbert action in the FG ambient space). The result was a sequence of seven equivalent action principles ending at the Einstein-Hilbert action (we review and extend that computation in Appendix~\ref{A}). Physically, the model corresponds to coupling a ``conformal geometry multiplet'' to a ``dilaton multiplet''.
Despite this nice physical interpretation, the model suffered a serious shortcoming; namely that one first solved the GJMS algebra problem, substituted the result into the dilaton multiplet action and then successively integrated out auxiliaries to reach the Einstein--Hilbert theory. Clearly this ignores backreaction, the missing ingredient being a master action describing the coupled conformal geometry--dilaton system. Our candidate Chern--Simons matrix model of quantum gravity provides a mechanism for solving this backreaction problem.

The model we propose is an infinite dimensional-matrix Chern--Simons theory where the matrices are the space of observables of a supersymmetric Hilbert space.
The differential of the model is the BRST operator corresponding to the~$\frak{sl}(2)$ Lie algebra cohomology differential. In fact this means that the model is the minimal
BV formulation of an underlying ``matrix'' model with a Chern--Simons BV action given by an AKSZ construction~\cite{AKSZ}. Our Article is structured as follows: In the next Section we give some further background details and state our model. In Section~\ref{R} we explain why this is a model of quantum gravity and how  the earlier backreaction problem is solved. In Section~\ref{O}, we focus on the model's linearization and gauge fixing; these are amenable to quantum mechanical path integral techniques. In the Conclusion we delineate various open problems and discuss the outlook for model building and a mathematical well-defined approach to quantum gravity. Appendix~\ref{A} reviews how gravity can be obtained by coupling conformal geometry moduli to a dilaton multiplet.

%
%
%
%
%
%
%
%
%
%
%
%
%
%
%
%
%
%
%
%

\section{The model}

The space of all GJMS algebras can be used to encode conformal geometries. Hence our first task is to develop a ``conformal geometry'' multiplet
and accompanying action principle whose solutions are GJMS algebras, this is done in Section~\ref{CGM}. This model has a large gauge invariance
which we handle using BV machinery in Section~\ref{BVAKSZ}. Thanks to the AKSZ construction, this model is governed by a Chern--Simons-type action.
To obtain a candidate quantum gravity model, the conformal geometry multiplet must still be coupled to scale. This is achieved in Section~\ref{Dilaton} by supersymmetrizing the BV extension of the theory; this  introduces a dilaton multiplet.

\subsection{Conformal geometry multiplet}\label{CGM}

The conformally improved scalar wave equation
$$
\Big[\, \Delta-\frac{d-2}{4(d-1)}\, R\, \Big]\varphi =0\, ,
$$
in~$d$-dimensions may be recast as triple of equations in a~$(d+2)$-dimensional FG ambient space\footnote{See~\cite{GJMS}; to explicitly verify  this, solve the first equation by writing~$\Phi = \delta(X^2)\varphi$ so that the second equation implies that~$\varphi$ is a weight~$1-\frac d2$ conformal density in~$d$-dimensions. The ambient Laplace equation then descends to the conformally improved scalar wave equation. This method underlies the standard construction of irreducible representations from wave equations~\cite{Helgason}, see~\cite{Gover:2009vc} for an account
of how it extends to curved spaces and tractor calculus. It has also been extensively employed in the $2T$ literature~\cite{Bars:2006dy}.}
$$
X^2 \Phi = 0 \, ,\qquad \Big[\nabla_X+\frac{d+2}{2}\Big]\, \Phi = 0\, , \qquad \Delta \Phi=0\, ,
$$
with metric obeying the closed homothety condition~\eqn{homothety}. These three---scalar singleton---conditions are exactly those imposed by the Dirac quantization of the GJMS algebra~\eqn{GJMS}. This suggests (see~\cite{BarsSol}), that gravity can be studied by considering the space of  operators obeying an~$\frak{sl}(2)=\frak{so}(2,1)$ algebra. More specifically we propose, as suggested by~\cite{Bars,BarsRey},
considering quantum mechanical observables subject to:
\begin{equation}\label{eoms3}
{}[{\bm Q}_a,{\bm Q}_b]=\varepsilon_{ab}{}^{c} \bm Q_{c}\, ,
\end{equation}
where~$a=(\pm,0)$ and indices are raised and lowered with the~$\frak{so}(2,1)$ metric~$\eta_{ab}$ where~$\eta_{+-}=1=\eta_{00}$.
We call the observables~$\bm Q^a$ the ``conformal multiplet'' and the solution space of~\eqn{eoms3} ``conformal moduli''.
A non-commutative field theory action principle underlying these equations of motion based on a star product was given in~\cite{BarsRey}.
However, rather than work with star products, since we ultimately are interested in diffeomorphism invariant systems, it is better to work directly with operators;\footnote{Albert Schwarz--private communication. This has also been used in~\cite{Bars:2002nu}.} thus we view the
${\bm Q}_a$ as infinite dimensional matrices with a trace operation given by any complete set of states on the Hilbert space~$\cal H$ so that~$${\rm Tr}_{\cal H} \bm O:=\sum_\alpha \langle \alpha| \bm O|\alpha\rangle\, .$$
Note, that for our purposes ${\cal H}$ is not a positive definite Hilbert space, but instead an indefinite relativistic Hilbert space of an ambient space with two timelike directions.
In these terms, the action principle is simply\footnote{In~\cite{BarsRey} an additional $U(1)$ observable is added to the $\frak{sp}(2)$ triplet to handle the dilaton. As we show later, this can be achieved, while at the same time solving the back reaction problem, by instead enlarging the quantum mechanical Hilbert space. Note also that large $N$ matrix models reminiscent of the above have been studied in~\cite{Smolin}.}
\begin{equation}
S={\rm Tr}_{\cal H}\Big[\,
\frac12\, {\bm Q}_a {\bm Q}^a+
\frac13 \, \epsilon^{abc}{\bm Q_a}{\bm Q_b}{\bm Q_c}\,
\Big]\, .
\end{equation}
This model has a large gauge invariance
$$
\bm Q_a\sim \bm Q_a + [\bm Q_a, \bm \epsilon]\, ,
$$
for any operator~$\bm \epsilon$. 
The equations of motions~\eqn{eoms3} can be solved by fixing most of this gauge freedom, leaving residual symmetries corresponding to diffeomorphisms of the ambient manifold and~$\frak{so}(1,1)$ Maxwell transformations. These invariances are in fact quite propitious in a conformal geometry situation; indeed the remaining conformal geometry moduli are then an ambient FG metric and Maxwell field.

\subsection{Minimal BV and AKSZ formulation}\label{BVAKSZ}

Returning to an off-shell setting, to handle the model's gauge invariance, we enlarge the ``field space''\footnote{Strictly speaking, because spacetime is emergent in this model, the dynamical variables are operators not fields, nonetheless we shall henceforth employ this abuse of language.} to its minimal BV content: fields~$\Phi^\alpha$ (= ghosts, gauge fields) and corresponding antifields~$\Phi^*_\alpha$; their names and Grassmann parities  are given as follows:$$
\begin{array}{c|
|c|c|c|c}
\mbox{field}&\ \bm C\ &\ \bm Q_a\ &\bm Q^*_{ab}:=\frac12\epsilon_{abc}\bm Q^{*c}&\bm C^*_{abc}:=\frac1{3!}\epsilon_{abc} \bm C^*\\[.5mm]\hline
\mbox{parity}&-&+&-&+\\[.5mm]
\end{array}
$$
By introducing odd coordinates~$c^a$, the above can be neatly packaged in a single ``AKSZ'' field~\cite{AKSZ} (see also~\cite{BarsBRST})
$$
\bm A:={\bm C}+{\bm Q}_a c^a +  {\bm Q}_{ab}^* c^a c^b +  {\bm C}^*_{abc} c^a c^b c^c\, .
$$
The minimal BV action is a sum of the classical action plus antifields multiplied by BRST variations of the fields:~$S_{\rm cl}[\Phi^\alpha]+\Phi^*_\alpha\,  \delta_{\rm BRST}\Phi^\alpha$. The field space~$(\Phi^\alpha,\Phi^*_\alpha)$ is a~$Q$-manifold~\cite{SchwarzQ} endowed with an odd symplectic structure (and hence a BV bracket) and a nilpotent vector field (generated by the BV action and BV bracket~$(S_{BV},\, \cdot\, )_{BV}$).
The quantum action is given by the BV action along a Lagrangian submanifold of this~$Q$-manifold.
The geometry of such~$Q$-manifolds was studied in~\cite{AKSZ} who noted that the minimal BV action for Chern--Simons theories was a ``master'' Chern--Simons theory. That situation applies here, where the minimal BV action is simply\footnote{Performing the Grassmann integration this action can equivalently be written
$$
S={\rm Tr}_{\cal H} \Big[\,
\frac12\, {\bm Q}_a {\bm Q}^a+
\frac13 \, \epsilon^{abc}{\bm Q}_a{\bm Q}_b{\bm Q}_c
-{\bm Q}^{*a}\big[{\bm C},{\bm Q}_a\big]
 -\frac12{\bm C}^*\big\{ {\bm C},{\bm C}\big\}\Big]
\, ,$$
which exhibits the BRST transformations of the fields and ghosts.
}
\begin{equation}\label{QQQ}
S={\rm Tr}_{\cal H} \int d^3c \, \Big[\, \frac12 {\bm A} \, {\frak d}  {\bm A} + \frac13 {\bm A}^3\Big]\, .
\end{equation}
In this formula, the nilpotent operator
$$
\frak d:=\frac 12\,  c^b c^a \epsilon_{ab}{}^c\frac{\partial}{\partial c^c}
$$
is the BRST operator/differential of the Lie algebra cohomology~$H^*(\frak{so}(2,1))$.\footnote{This theory was developed over sixty years ago in the mathematics literature, the book~\cite{Fuchs} gives an excellent account, for a computation of  the cohomology of~$\frak d$ in a physics context, see the Appendix of~\cite{BarsY}.}
The mechanics of the analogous three dimensional Chern--Simons computations carries over to show that the above action (i) enjoys the gauge invariance
$$
{\bm A}\sim {\bm A}+ {\frak d}_{\bm A} \bm {\varepsilon}\, ,
$$
with operator valued Grassmann-even gauge parameters ~$\bm {\varepsilon}=\bm {\varepsilon}(c)$ and covariant derivative
$$
{\frak d}_{\bm A} :={\frak d}+[{\bm A},\, . \, \}\, ,
$$
where~$[\, . \, , \, .  \}$ denotes a graded commutator; (ii) is extremal on flat connections given by the zero curvature condition
$$
{\bm F}_{\bm A}:={\frak d}_{\!\bm A}^{\, 2} = {\frak d} {\bm A} + {\bm A}^2 = 0\, ;
$$
and (iii) linearized about a solution~$\bar {\bm A}$, fluctuations~$\bm a:=\bm A-\bar {\bm A}$ obey the parallel condition
$$
\frak d_{{\bar{\bm A}}} \bm a=0\, ,
$$
modulo linearized gauge transformations~$\bm a\sim \bm a + \frak d_{{\bar{\bm A}}} \bm \varepsilon$. Note that the zero curvature condition implies that $\frak d_{{\bar{\bm A}}}$ is nilpotent: 
\begin{center}
\shabox{
$\frak d_{{\bar{\bm A}}}^{2}=0$}
\end{center}
\vspace{-12mm}
\begin{equation}\label{gruyere}\phantom{stuff}\end{equation}
\vspace{1mm}

\noindent
so this system is cohomological, and in fact amenable to a quantum mechanical analysis; see Section~\ref{O}. 

\subsection{Dilaton coupling}\label{Dilaton}

Up to this point, we have only discussed the model describing the conformal geometry moduli. However, having expressed this in its BV form,
coupling to a dilaton multiplet is simple. For that, we  supersymmetrize  the Chern--Simons algebra of quantum mechanical observables.
In the BV formalism every field has a corresponding antifield of opposite Grassmann parity so supersymmetrizing the BV description of the model and viewing the superpartners of antifields as further fields avoids introducing physical superpartners.
We introduce internal Grassmann coordinates~$(\gamma,\overline\gamma)$ and replace all fields by superfields~$(\Phi^\alpha,\Phi_\alpha^*)\mapsto
(\Phi^\alpha(\gamma),\Phi_\alpha^*(\gamma))$. The Hilbert space trace becomes a supertrace~${\rm Str}_{\cal H}:={\rm Tr}_{\cal H}\, \int d^2\gamma$ (the reader should not confuse the slightly longer bar notation for complex conjugation with that for background solutions).
The action is
\begin{center}
\shabox{$ \displaystyle S={\rm Str}_{\cal H} \int d^3c \, \Big[\, \frac12 {\cal A} \, {\frak d}  {\cal  A} + \frac13 {\cal A}^3\Big] $}
\end{center}
\vspace{-13mm}
\begin{equation}\label{swisscheese}\phantom{stuff}\end{equation}
\vspace{3mm}

\noindent
In the following Section, we argue that this theory is a model for quantum gravity. To that end, we record a few basic facts about the theory:
(i) It enjoys a gauge symmetry
$$
{\cal A}\sim {\cal A}+\frak d_{\cal A} \cal E\, ;
$$
(ii) the action is extremal on flat connections obeying the zero curvature condition
$$
{\cal F}_{\cal A}:={\frak d}_{\!\cal A}^{\, 2} = {\frak d} {\cal A} + {\cal A}^2 = 0\, ;
$$
and (iii) its linearization proceeds exactly as discussed above. To exhibit the minimal BV nature of the action
we can perform the integration over the Grassmann coordinates~$c^a$ and find
$$
S={\rm Str}_{\cal H} \, \Big[
\frac12\, {\cal Q}_a {\cal Q}^a+
\frac13 \, \epsilon^{abc}{\cal Q}_a{\cal Q}_b{\cal Q}_c
-{\cal Q}^{*a}\big[{\cal C},{\cal Q}_a\big]
-\frac12\, {\cal C}^*\big\{ {\cal C},{\cal C}\big\}\Big]\, ,
$$
where we have defined
$$
{\cal A}:={\cal C}+{\cal Q}_a c^a + \frac12\, \epsilon_{abc} {\cal Q}^{*a} c^b c^c + \frac1{3!}\, {\cal C}^*\epsilon_{abc} c^a c^b c^c\, .
$$
Alternatively we can perform the integral  over the internal Grassmann coordinates~$(\gamma,\overline\gamma)$ and find
\begin{equation}\label{bluecheese}
S= {\rm Tr}_{\cal H} \int d^3c\ \Big\{{\bm \chi}\,  {\bm F}_{\bm A}+ \overline{\bm \Psi}\,  {\frak d}_{\bm A} {\bm \Psi}\Big\}\, ,
\end{equation}
where the superfield~$\cal A$ has the expansion
$
{{\cal A}}(c,\gamma):={\bm A}+ \overline{\bm \Psi} \gamma+ {\bm \Psi} \overline\gamma+ {\bm \chi} \gamma\overline\gamma
$.

\section{Quantum Gravity}\label{R}

We now analyze how the functional integral, weighted by the action~\eqn{swisscheese},  produces  a  sum over conformal geometries, and thus  models quantum gravity.
First remember that there are two equivalent ways of presenting the action (obtained by integrating explicitly over the~$c^a$ or~$(\gamma, \overline\gamma)$ Grassmann coordinates) which  manifest either the minimal BV structure or  the dilaton-conformal geometry coupling respectively. We begin with action in the form~\eqn{bluecheese} because it manifests
the dilaton-conformal geometry coupling and thus consider the (Euclidean\footnote{For the formal computations performed here, we could equally well consider a Lorenztian path integral.}) functional integral:
\begin{equation}\label{PartitionFunction}
Z=\int [D{\bm\chi}][D\overline{\bm \Psi}][D{\bm \Psi}][D{\bm A}]\,{\rm exp}\left\{-{\rm Tr}_{\cal H} \int d^3c\ \Big[{\bm \chi}\,  {\bm F}_{\bm A}+\overline{\bm \Psi}\,  {\frak d}_{\bm A} {\bm \Psi}\Big]\right\}\;. 
\end{equation}
Note that  this functional integration over operator-valued fields could also be viewed as an (infinite dimensional) integral over matrix elements of the operators themselves, or alternatively as a path integral over an infinite tower of ambient space tensor fields  which arise by expanding operators in powers of~$\nabla_M$. 
Since the action we integrate over is already of BV-type, we do not need any further extension of the field space to  deal with gauge symmetries. Moreover, in BV perturbation theory one can use as propagator a partial inverse of the BV kinetic operator~\cite{SchwarzFix}.\footnote{By partial inverse we mean that the propagator~$\bm G$ is an inverse under an  adjoint action, \emph{i.e.}~$\big[{\bm K, \,\bm G}\big\}=\bm 1$, where~$\bm K$ is the kinetic quadratic form. The partial inverse is determined only up to an equivalence class that reflects the arbitrary choice of a Lagrangian submanifold.} By doing so one neither needs to introduce a non minimal BV sector, nor  choose a gauge fixing fermion: indeed the choice of propagator is equivalent to a choice of gauge fixing fermion in the usual setting.

Having discussed how the model's gauge symmetries are correctly dealt with at the quantum level, we are now ready to perform some formal manipulations on the path integral~\eqref{PartitionFunction}.
To begin with, we notice that integrating over~$\bm\chi$ imposes a zero curvature condition; in fact it precisely solves the backreaction problem described in the  Introduction:
\begin{equation}
\begin{split}
\int [D{\bm \chi}]\,
\exp\Big\{-{\rm Tr}_{\cal H} \int d^3c\,{\bm \chi}\,  {\bm F}_{\bm A}\Big\}&=\delta\left({\bm F}_{\bm A}\right)\\&=\sum_{\bar{\bm A}}\delta\left({\bm A}-{\bar{\bm A}}\right)\left[\det\left.\frac{\delta{\bm F_{\bm A}}}{\delta{\bm A}}\right\rvert_{{\bm A}={\bar{\bm A}}}\right]^{-1}\\[2mm]
&=\sum_{\bar{\bm A}}\delta\left({\bm A}-{\bar{\bm A}}\right)\left(\det{\frak d}_{\bar{\bm A}}\right)^{-1}\;,
\end{split}
\end{equation}
where the
sum
over flat connections is generically a path integral possibly combined with a sum over distinct topological sectors.
This  result can be inserted in~\eqref{PartitionFunction}, which, remembering that  $(\overline{\bm \Psi}, {\bm\Psi})$ are Grassmann even, allows the integrations over  remaining fields to be performed:
\begin{center}
\shabox{$\displaystyle Z=\sum_{\{ \bar{\bm A}|\bm F_{\bar{\bm A}}=0\}}\left[\det{\frak d}_{\bar{\bm A}}\right]^{-2}\;$}
\end{center}
\vspace{-13mm}
\begin{equation}\label{sumflatA0}\phantom{stuff}\end{equation}
\vspace{3mm}

\noindent
where~${\frak d}_{\bar{\bm A}} :={\frak d}+[{\bar{\bm A}},\, . \, \}$ acts in the (operator) adjoint representation.
At this point, the functional determinant~$\det{\frak d}_{\bar{\bm A}}$ could be computed in BV perturbation theory (see Section~\ref{O}) but for now we are more interested in relating this result to quantum gravity. The fact that the path integral localizes over flat connections is not so surprising from a Chern--Simons viewpoint; this   hints that   quantum gravity partition functions can be better mathematically defined as infinite dimensional matrix models. A first step in that direction is to show that the partition function~\eqref{sumflatA0} includes an integral over conformal classes of metrics.
To see this we recall that the superfield~$\bm A(c)$ contains the gauge fields~$\bm Q_a$, together with their ghosts and antifields. Hence, the integration over flat connections contains an integral over conformal geometry moduli~$\bar{\bm Q}_a$ solving\footnote{Strictly, the flatness condition $\bm F_{\bm A}=0$ amounts, in the~$\bm Q_a$ sector, to
$
[\bar{\bm Q}_a,\bar{\bm Q}_b]-\epsilon_{abc}{\bar{\bm Q}}_c=\epsilon_{abc}\{\bar{\bm C},\bar{\bm  Q}^{*c}\}\, .
$
The right hand side of this is BRST exact; we omit it because we only turn on the conformal geometry moduli~$\bm Q_a$.}~\eqn{eoms3}.
Remarkably, this gives a solution to the backreaction problem, since the supersymmetric coupling to the dilaton multiplet still implements the~$\frak{sp}(2)$ algebra condition~\eqref{eoms3} governing the conformal geometry moduli space.
Solutions to~\eqn{eoms3} which solve the flatness condition activating only the~$\bar{\bm Q}_a$ moduli only were given in~\cite{Bonezzi} and are reproduced in Appendix~\ref{A}. They depend on an ambient FG metric and Maxwell field $(g_{MN},A_M)$.
%
Thus we see  that the formal sum over flat connections in~\eqn{sumflatA0} includes a path integral over conformal geometry moduli
$$
\sum_{\bar{\bm A}}\  \supset\  \int [Dg_{MN}][DA_M]   \;.
$$
For the moment we refrain  from trying to analyze the whole moduli space coming from the flatness condition~$\bm F_{\bar{\bm A}}=0$,
for our current purposes it suffices that conformal geometries are included in this space; we will return to this issue in our Conclusions.
Also, the appearance of an integral over conformal geometries alone is not enough to show that we are dealing with a model of quantum gravity.
We still need to show that the quantum measure, at  least in a ``diagonal limit'', is governed by the exponential of the Einstein--Hilbert action.
In fact, in~\cite{Bonezzi} it was shown (see also  Appendix~\ref{A}) that classical gravity arises when coupling conformal geometry moduli~$\bm Q_a$ to scale, \emph{i.e.} a  dilaton multiplet. In the framework of \cite{BarsBRST}, the Einstein-Hilbert action arose from a BRST-type lagrangian of the form
\begin{equation}\label{oldgravity}
S_{\rm gravity}=\int_{\tilde M}\,\Psi^a\,{\bar{\bm Q}}_a\Psi\;,
\end{equation}
where the conformal geometry moduli~${\bar{\bm Q}}_a$ are given explicitly in~\eqref{solns}, and~$(\Psi^a, \Psi)$ are ambient fields ({\it not} operators). In our present context the dilaton multiplet~$(\overline{\bm \Psi},\bm\Psi)$ is  also operator-valued, being on the same footing as the conformal geometry multiplet, and  consists of a minimal BV ``field'' content:
$$
{\bm \Psi}(c)={\bm \psi}+c^a\,{\bm \psi}^*_a+\frac12\,c^ac^b\,\epsilon_{abc}\,{\bm \psi}^c+\frac{1}{3!}\,c^ac^bc^c\epsilon_{abc}\,{\bm \psi^*}\;.
$$
 Grassmann parities are  given by 
\begin{align*}
\begin{array}{c||c|c|c|c}
\mbox{field} & {\bm \psi} & {\bm \psi}^*_a & {\bm \psi}^a & {\bm \psi}^*\\\hline
\mbox{parity} & + & - & + & -\\
\end{array}
\end{align*}
and the bar involution is defined as 
\begin{align*}
\overline{{\bm \Psi}}(c)=\overline{\bm \psi}+c^a\,\overline{{\bm \psi}}^*_a+\frac12\,c^ac^b\,\epsilon_{abc}\,\overline{\bm \psi}^c+\frac{1}{3!}\,c^ac^bc^c\epsilon_{abc}\,\overline{\bm \psi}^*
\end{align*}
The relevant interaction comes from the~$\overline{\bm \Psi}\,  {\frak d}_{\bm A} {\bm \Psi}$ part of the Lagrangian:
$$
{\rm Tr}_{\cal H}\,\Big[\, \overline{\bm\psi}{}^a\, [{\bm Q}_a,{\bm\psi}]+\overline{\bm\psi}\ [{\bm Q}_a,{\bm\psi}^a]\, \Big]\;.
$$
The coupling~\eqn{oldgravity} is in fact hidden in the above. To uncover it, we consider the diagonal limit where the operators~${\bm\psi}^a$ and~$\bm \psi$ are pure states up to a phase (so no sum over~$a$ in the following):
$$
{\bm\psi}_{\rm pure}=z\ketbra{\psi}{\psi}\;,\quad {\bm\psi}^a_{\rm pure}=w\ketbra{\psi^a}{\psi^a}\;,
$$
with~$z,w\in\mathbb{C}$
so that
$$
{\rm Tr}_{\cal H}\,\big[\overline{\bm\psi}^a[{\bm Q}_a,{\bm\psi}]+\overline{\bm\psi}[{\bm Q}_a,{\bm\psi}^a]\big] =2{\rm Re}\,\int\bar \Psi^{a}\bm Q_a \Psi
$$
where the ambient (Schr\"odinger representation) fields~$\Psi:=\Psi(y)=\braket{y}{\psi}$ and~$\Psi^a:=\Psi^a(y)=(w\bar z-z\bar w) \braket{\psi^a}{\psi}\,\braket{y}{\psi^a}$ (for~$y\in\tilde M$).
%
This precisely recovers a complexified version of the Lagrangian~\eqref{oldgravity}. We analyze this in detail in the Appendix and find a nonlinear sigma model coupled to gravity. Hence,  our model
gives a candidate theory of quantum gravity in the sense that 
$$
Z=\int [Dg\cdots] \exp\{-S_{\rm EH}+\cdots\}\, ,
$$
where the dots indicate corrections to an integration over metrics weighted by the exponential of Einstein--Hilbert action over both of which we do not yet have full control, due to our lack of understanding of the full moduli space of flat connections and the determinant~$\det{\frak d}_{\bar{\bm A}}$. These are in principle calculable. In the next Section we sketch approaches for handling the determinant.

\section{Effective Actions}\label{O} 
The expression appearing inside the sum~\eqref{sumflatA0}, for~${\bar{\bm A}}$ fixed, can be viewed as (the exponential of) a field theory one-loop effective action $\Gamma(\bar{\bm A})$ which can be handled  using the BV perturbative strategy devised in~\cite{SchwarzFix}. Indeed the {\it na\"ive} determinant in~\eqref{sumflatA0} is ill-defined as the Grassmann operator ${\frak d}_{\bar{\bm A}}$ 
 has zero modes due to the nilpotency condition~\eqn{gruyere} responsible for the 
 linearized gauge symmetry~$\delta {\bm a} ={\frak d}_{\bar{\bm A}}\bm \varepsilon$. It is thus propitious to treat 
 ${\frak d}_{\bar{\bm A}}$ as the BRST operator of an underlying quantum mechanical model. Focusing on backgrounds $\bar{\bm A}=c^a \bar{\bm Q}_a$ where only the 
 conformal geometry moduli backgrounds are turned on\footnote{Observe that the zero-curvature solution~\eqref{solns} is  not  pure gauge: $\bar {\bm A} = \bar {\bm Q}_a c^a$ has ghost number one, whereas nonvanishing terms in a pure gauge solution~$e^{-{\bm \lambda}} {\frak d} e^{{\bm \lambda}}$ have, at least, ghost number two. Hence~\eqref{solns} is a cohomologically non-trivial solution. Note also that more general backgrounds can also be analyzed by similar methods.}
we can rewrite it as
\begin{align*}
{\frak d}_{\bar{\bm Q}} := {\frak d}+[{\bar{\bm Q}_ac^a},\, . \, \} = {\frak d} + c^a [\bm{\bar Q_a},\, . \, \ ]  = c^a\Big (  {\bf D}_a+{\bf d}_a\Big)\, ,
\end{align*}
 where 
 \begin{align*}
 {\bf D}_a := [\bar {\bm Q}_a,\, . \, \ ]\, ,\qquad 
{\bf d}_a := \frac12\epsilon_{ba}{}^c c^b \frac{\partial}{\partial c^c}\, .
\end{align*}   
We can similarly construct a nilpotent antiBRST-like operator 
\begin{align}\nonumber
{\bm\delta}_{\bar{\bm Q}} = \frac{\partial}{\partial c^{a}}\, \eta^{ab}\Big (  {\bf D}_b- {\bf d}_b\Big)\, .
\end{align}
The latter allows us to partially invert the operator~${\frak d}_{\bar{\bm Q}}$ because
\begin{align*}
\big\{ {\frak d}_{\bar {\bm Q}},{\bm \delta}_{\bar{\bm Q}} \big \} = {\bm \Delta}_{\bar{\bm Q}}\neq 0\,,
\end{align*}
implies
\begin{align}\nonumber
 \left\{ {\frak d}_{\bar {\bm Q}},\frac{{\bm \delta}_{\bar{\bm Q}}}{{\bm \Delta}_{\bar{\bm Q}}} \right\} = 1\, . 
\end{align}
In the above we have
\begin{align}\nonumber
{\bm \Delta}_{\bar{\bm Q}}  = {\bf D}^2 +\frac12 {\bf N} ({\bf N}-3) \, ,\qquad {\bf N} := c^a\frac{\partial}{\partial c_a}
\end{align} 
which are (quantum mechanically) a central Hamiltonian and ghost number operator:
$$[{\frak d}_{\bar{\bm Q}},{\bm \Delta}_{\bar{\bm Q}}]=0=[{\bm \delta}_{\bar{\bm Q}},{\bm \Delta}_{\bar{\bm Q}}]\, ,\quad [{\bf N},{\frak d}_{\bar{\bm Q}}]={\frak d}_{\bar{\bm Q}}\, ,
\quad [{\bf N},{\bm \delta}_{\bar{\bm Q}}]=-{\bm \delta}_{\bar{\bm Q}}\, .$$
The inverse~$\frac{{\bm \delta}_{\bar{\bm Q}}}{{\bm \Delta}_{\bar{\bm Q}}} $ amounts to a Dirac-type propagator  in the presence of an external ``field" ~$\bar {\bm Q}$. Therefore, in Feynman diagram notation it corresponds to a sum of infinitely many graphs:
\begin{align*}
\frac{{\bm \delta}_{\bar{\bm Q}}}{{\bm \Delta}_{\bar{\bm Q}}}\, = \, 
\parbox{60pt}{
     \begin{fmfgraph}(60,60)
       \fmfleft{i} \fmfright{o} 
       \fmf{plain}{i,o} 
   \end{fmfgraph}}\ + \
   \parbox{60pt}{
     \begin{fmfgraph}(60,60)
       \fmfleft{i} \fmfright{o} \fmftop{t} 
       \fmf{plain}{i,v} \fmf{plain}{v,o} 
       \fmffreeze
        \fmf{photon}{t,v}
   \end{fmfgraph}} 
   \ + \
   \parbox{60pt}{
     \begin{fmfgraph}(60,60)
       \fmfleft{i} \fmfright{o} \fmftop{t1,t2} 
       \fmf{plain}{i,v1,v2,o} 
       \fmffreeze
        \fmf{photon}{t1,v1} \fmf{photon}{t2,v2}
   \end{fmfgraph}}
   \ + \
   \parbox{60pt}{
     \begin{fmfgraph}(60,60)
       \fmfleft{i} \fmfright{o} \fmftop{t1,t2} 
       \fmf{plain}{i,v1,v2,o} 
       \fmffreeze
        \fmf{photon}{t1,v2} \fmf{photon}{t2,v1}
   \end{fmfgraph}}\ +\
    \cdots\, ,
 \end{align*}



\noindent with an arbitrary number of insertions of the external ``potential''. One way to represent it is by using the worldline formalism: firstly one exponentiates the propagator using  the superSchwinger trick  
\begin{align*}
\frac{{\bm \delta}_{\bar{\bm Q}}}{{\bm\Delta}_{\bar{\bm Q}}} 
= \int_0^\infty dT \int d\Theta ~e^{-T{\bm \Delta}_{\bar{\bm Q}} -{\bm \delta}_{\bar{\bm Q}} \Theta}=:\bm P\, ,
\end{align*} 
where~$\Theta$ is a Grassmann variable, and then treats~${\bm \Delta}_{\bar{\bm Q}}$ and ${\bm \delta}_{\bar{\bm Q}}$ as operators in  single particle quantum mechanics. In fact, thanks to their centrality and nilpotency properties, they can be interpreted as a pair of abelian, first class constraints. Representing the operator-valued integrand of the above as a worldline path integral in a (super)phase space~$(Z,\omega)$,
schematically one can write the Greens function for the partial propagator~$\bm P$ as
\begin{align}
\label{eq:world-prop}
P\big(z_{\rm i},z_{\rm f}\big) =\int_0^\infty dT \int d\Theta 
\int_{z_{\rm i}}^{z_{\rm f}} [dz]~\exp\left\{\int_{\rm i}^{\rm f} \Big[\, \theta -\big(T\Delta(z) +\delta(z) \Theta\big)d\tau\Big]\right\}\, ,
\end{align}
where~$z\in Z$ and we have locally integrated the symplectic form $\omega=d\theta$ to a symplectic current $\theta$. The operators ${\bm \Delta}_{\bar{\bm Q}}$ and 
${\bm \delta}_{\bar{\bm Q}}$ are here replaced by their corresponding classical Hamiltonians~$\Delta(z)$ and~$\delta(z)$.
To obtain the effective action, one ``glues"  together the propagator end-points and traces over them. 
This model amounts to the minimal quantum mechanical BV treatment of the linearization of the GJMS algebra equations~\eqn{eoms3}.

In general, understanding how to correctly glue propagator endpoints to obtain an effective action is rather intricate.
A way to circumvent those difficulties   is to notice that the above propagator can be thought of as a gauge-fixed worldline path integral for a locally (super)symmetric particle action, where the superSchwinger times are nothing but  moduli for particle gauge fields and  the first class constraints generate gauge transformations (reparameterization and local supersymmetry) for the dynamical worldline fields~$z$. Therefore the expression~\eqref{eq:world-prop} can be written as
\begin{align}\nonumber
P(z_{\rm i},z_{\rm f}) =\int_{\ell} \frac{[dz] [de]}{\rm Vol(gauge)}\,  e^{-S[z,e]}\, ,
\end{align}   
where~$\ell$ indicates that we are computing a path integral with a ``line" topology ({\it i.e.}, fixed boundary conditions) and~$e$ collectively denotes the particle gauge fields of the  worldline action~$S[z,e]$ whose gauge fixing on the line leads to the action in~\eqref{eq:world-prop}. Finally one  obtains the effective action by taking the same path integral but with a circle topology--{\it i.e.} (anti)periodic boundary conditions:  
\begin{align}\nonumber
\Gamma[{\bar{\bm Q}}] = \int_{S^1} \frac{[dz] [de]}{\rm Vol(gauge)}\, e^{-S[z,e]} \, .
\end{align}
The above particle path integral can be computed by gauge fixing the worldline action using Hamiltonian BRST methods:  One adds (further non-minimal) ghosts~${\rm c}$ and ghost momenta~${\rm \pi}$ for all gauge symmetries and develops an extended BRST operator as a graded sum (in the ghost momenta)~$\Omega= \sum_p {\Omega}_p$, so that the quantum Hamiltonian becomes~$H_{\rm qu} =H_{\rm BRST}+\{K,\Omega\}$ where~$H_{\rm BRST}$ is a BRST-invariant Hamiltonian and~$K$ a gauge-fixing fermion. If the particle action is worldline-diffeomorphism invariant the Hamiltonian itself enters as a constraint ({\it i.e.}, a local-symmetry generator) and we can set~$H_{\rm BRST}=0$. This procedure leaves a set of modular parameters~$t_k$ that must be integrated over a fundamental domain (FD); they parametrize gauge-inequivalent configurations. Hence,
\begin{align}\nonumber
\Gamma[{\bar{\bm Q}}] &=  \int_{\rm FD} \prod_k dt_k \int_{S^1} [dz] [d{\rm c}] [d{\rm \pi}]~e^{-S_{\rm qu} [z,\hat e(t),{\rm c},{\rm \pi}]}
=\sum\ \parbox{50pt}{
     \begin{fmfgraph}(70,70)
       \fmfleft{i} \fmfright{o} 
       \fmftop{t1,t2,t3,t4} \fmfbottom{b1,b2,b3,b4} 
       \fmf{plain}{l,v1,v2,v3,r} \fmf{plain}{l,u1,u2,u3,r}
       \fmf{photon}{i,l} 
        \fmf{photon}{t1,v1} \fmf{photon}{t2,v2} \fmf{photon}{t3,v2} \fmf{photon}{t4,v3}
         \fmf{photon}{r,o} 
        \fmf{photon}{b1,u1} \fmf{photon}{b2,u2} \fmf{photon}{b3,u2}  \fmf{photon}{b4,u3} 
   \end{fmfgraph}}\qquad\, ,
\end{align}   
where~$\hat e(t)$ are the fixed gauge fields and~$S_{\rm qu} [z,\hat e(t),{\rm c},{\rm \pi}] = \int_0^1(\theta_{\rm gh} -H_{\rm qu}d\tau )$ where $\theta_{\rm gh}$ is the ghost-extended symplectic current. As depicted the one-loop effective action describes a sum of one-particle irreducible diagrams with insertions of external fields.
We plan to report on this computation in a future publication~\cite{BOW}.
}
\end{fmffile}

\section{Conclusions and Outlook}\label{Outlook}

In this paper we have proposed a Chern--Simons matrix model for quantum gravity. Its input data is only a Hilbert space whose observables play the {\it r\^ole}
of the space of matrices integrated over, or in other words the model is defined by a choice of quantum mechanics. For the choice given by the quantum mechanics of a~$(d+2)$-dimensional ambient space, we found that the model can be written as a sum of~$d$-dimensional causal stuctures plus further moduli determined by a certain zero curvature condition. We showed that the leading path integral measure was the exponential of the Einstein--Hilbert action. 
Spacetime is emergent\footnote{In the $2T$ shadow picture of~\cite{Bars}, one could  hope that a landscape of dual spacetimes could  emerge from these shadows.} in this model:
 the ambient space~$\tilde M$---and hence spacetime $M$ equipped with a causal structure---arises from a dual pair construction~\eqn{dual}.

There are  many open questions. The situation is somewhat reminiscent of the early development of String Theory: First we need to know what propagating degrees of freedom (DoF) the model describes. The problem here is that simpler models involving only metric degrees of freedom could only be treated by ignoring  backreaction (see Appendix~\ref{A}). To determine the DoF of the full  model requires us to (i) solve the zero curvature condition $\bm F_{\bar{\bm A}}=0$;~and (ii) compute the determinant~$\det \frak{d}_{\bar{\bm A}}$. It seems rather unlikely that this yields only the metric fluctuations and Einstein--Hilbert dynamics that we found by specializing to pure states. However, just as is the case for the infinite tower of (gapped) massive string states, additional propagating modes and accompanying dynamics could well be a virtue. Indeed, one can even speculate that the finiteness properties of the underlying matrix model may be better than that of an integral over metrics. Moreover, one might try to regulate the sum over quantum mechanical observables by hermitean matrices, in which case a slew of random matrix model techniques could be bought to bear on the problem; indeed the model itself is structurally very close\footnote{Recently~\cite{BarsSFT} appeared which actually uses methods similar to proposed here to analyze String Field Theory itself.} to String Field Theory~\cite{SFT} which has been amenable to a matrix model approach~\cite{Marino}.
In particular, we note that we need not require strict finiteness, but only renormalizability of the matrix model.

If it is truly the case that the model we have proposed is better defined than a path integral over metrics, then an urgent problem would be to study how to build models in this framework; in particular coupling to matter fields and their stress energy would be a pressing question. There is much work to be done here, since at present we have only a rudimentary understanding of how to couple the conformal geometry multiplet to a dilaton to yield gravity. Nonetheless, it is interesting to observe that supersymmetry already played a part here, without necessarily implying that elementary particles come in bose-fermi marriages. Instead, working in the BV formalism, adding additional supersymmetry in fact just gave additional bose partners for bose fields.

Another interesting feature of the model is that spacetime plays a rather secondary {\it r\^ole} because it only enters through a particular Schr\"odinger representation of the input Hilbert space. If our proposed model is to be a useful formulation of quantum gravity, it ought be able to see the types of dualities present in leading approaches such as string theory. That this would require a model where spacetime is an emergent quantity is perhaps not surprising.

\section*{Acknowledgements} It is a pleasure to thank Itzhak Bars and Albert Schwarz for discussions. The authors thank  UCMEXUS-CONACYT grant CN-12-564 for partial support. 

\appendix

\section{The Gravity Sector}\label{A}

Here we show why the complexified version of the action~\eqn{oldgravity} is equivalent to Einstein--Hilbert gravity. This account follows directly the one given in~\cite{Bonezzi}. The starting point is the model
$$
S_{\rm gravity}[\bar{\bm Q}_a,\Psi^a,\Psi]=\int_{\tilde M} \Big[\, \overline \Psi^a \bar{\bm Q}_a \Psi + \overline \Psi\,  \bar{\bm Q}_a \Psi^a\Big]\, , 
$$ 
It is important to note that here~$(\Psi^a,\Psi)$ are fields on the ambient manifold while~$\bm Q_a$ are operators.
Varying~$\Psi^a$ imposes the triplet  of field equations
\begin{equation}\label{psiQpsi}
 \bar{ \bm Q}_a \Psi=0\, .
\end{equation}
The next ingredient is the on-shell conformal geometry multiplet
\begin{equation}\label{solns}
\bar{\bm Q}_+=\frac1{2\sqrt{2}}\, X^M g_{MN}X^N\, , \quad \bar{\bm Q}_0 = \frac12\left(\nabla_X(A) + \frac{d+2}{2}\right)\,  ,\quad
\bar{\bm Q}_{-}=-\frac1{2\sqrt{2}}\, \nabla_M(A)\, \nabla^M(A)\, ,
\end{equation}
where the ambient metric is the gradient of a homothety
$$
g_{MN}=\nabla_M X_N
$$
and the connection~$\nabla_M(A):=\nabla_M+A_M$ whose Maxwell curvature~$F_{MN}$ of the~$\frak{so}(1,1)$ gauge field~$A_M$ obeys
$$
X^M F_{MN}=0\, .
$$
In the above it is possible to add a higher spin branch to the the solutions by adding terms  $\Sigma + H\big(\nabla(A)\big)$ to 
the operator $\bar{\bm Q}_-$ where the scalar $\Sigma$ obeys $\nabla_X\Sigma = -2\Sigma$ and $H$ is an expansion in~$\nabla$ with coefficients of $\nabla^s$
obeying $(\nabla_X+2-s)H_{M_1\cdots M_s}=0=X^M H_{MM_2\ldots M_s}=0$ and $s\geq2$. This higher spin branch was first discovered in~\cite{BarsSol} appearing in the classical 
solution to~\eqn{eoms3} where quantum commutators were replaced by Poisson brackets. In~\cite{Bonezzi}, it was shown that at the quantum level, the higher spin branch can be gauged away, so long as\footnote{This point is perhaps slightly subtle: solutions with $\Sigma\neq 0$ and  $H\neq0$ are gauge equivalent
to solutions where {\it both} $\Sigma=0$ and $H=0$.}~$\Sigma\neq 0$. We proceed therefore, to analyze the case where the higher spin branch of solutions is absent.
Thus, evaluated on solutions~\eqn{solns} we have an action depending only on ambient fields (here we rescaled fields to normalize coefficients)
$$
S[g_{MN},A_M,\Psi,\Psi^a]={\rm Re}\int_{\tilde M}\Big[
\overline\Psi^+ X^2 \Psi + \overline\Psi^0 \big(\nabla_X(A)\, \Psi + \frac{d+2}{2}\, \Psi\big) + \overline\Psi^- \Delta(A)\, \Psi \big)
\Big] \, .
$$
This action enjoys residual gauge invariances
\begin{align}
A_M\, &\sim \hspace{-1.8cm}&A_M&+\nabla_M\alpha\, ,\nonumber\\[2mm]
 \Psi\ \ \,  &\sim\hspace{-1.8cm} &\Psi\ \ &  -\ \alpha \, \Psi\, ,\nn\\
\Psi^-\ &\sim \hspace{-1.8cm}&\Psi^-&+\alpha\, \Psi^--X^2\theta-\big(X^M\nabla(A)_M+\frac{d+2}{2}-2\big)\, \omega\, ,\nonumber\\
\Psi^0\ \,   &\sim\hspace{-1.8cm} & \Psi^0\, & +\alpha\, \Psi^0\, +X^2\lambda-\Delta(A)\, \omega-4\, \theta\, ,\nonumber\\
\Psi^+\ &\sim\hspace{-1.8cm} &\Psi^+ &+ \alpha\, \Psi^++\Delta(A)\, \theta +\big(X^M\nabla(A)_M+\frac{d+2}{2}+2\Big)\lambda\, .\nonumber
\end{align}
\vspace{-1.2cm}
\begin{equation}\label{ggauge}
\end{equation}
\noindent
Here the local  parameter $\alpha$ is real while $(\theta,\omega,\lambda)$ are complex.
Now we integrate out two of the Lagrange multipliers~$\Psi^+$ and~$\Psi^0$
which imposes
$$
\Psi=\delta(X^2) \phi\, ,\qquad
\phi\sim \phi+X^2 \chi\, ,
$$
as well as
$$
\Big(\nabla_X-2+X^M A_M +\frac{d+2}2 \Big)\phi=0\, .
$$
The Maxwell invariance of the model (with parameter~$\alpha$ in~\eqn{ggauge}) can be used to choose a gauge for the top slot~$X^MA_M=-w$ so this condition then implies that
$\phi$ is a conformal density of weight~$w-1+\frac{d}{2}$ on the conformal manifold~$M$. (Our final result will not depend on the choice of~$w$; note that in the ambient description, weights are given by the eigenvalue of~$\nabla_X$). There is still the freedom using the gauge parameter~$\omega$ to mostly gauge away~$\overline\Psi^-$ (this exhausts the  gauge transformations with parameter~$\omega$ save for~$\omega$ in the kernel of~$X^M \nabla_M(A) +\frac d2 -1$). Hence all that remains is the part $\overline\psi$ of~$\overline\Psi^-$ with weight~$-w-\frac d2 +1$
so the remaining fields and their  weights are now
 \begin{center}
 \begin{tabular}{c|cccc}
Field&$\overline\psi$&$\phi$&&$A_M$\\[1mm]\hline\\[-3mm]
Weight&$-w-\frac d 2 +1$&$w-\frac d 2 +1$&&$-1$
\end{tabular}
\end{center}
The action becomes (up to an unimportant normalization and integrations by parts ensuring no derivatives act on the delta function)
\begin{equation*}\label{aAction2}
S= {\rm Re}\int_{\tilde M}  \delta(X^2)\, {\cal I}\, ,\qquad {\cal I}:=  \overline{\phi \,} (\nabla^M-A^M)(\nabla_M-A_M) \psi,
\end{equation*}
Since the quantity~${\cal I}$ multiplying the delta function has definite weight and is defined up to the equivalence~\eqn{coneE}, it is  a  weight~$-d$ conformal density, and thus can be expressed in terms of tractors~\cite{BEG}:\be\nn
{\cal I}\sim \overline{\phi\,} \Big[ \frac 1 w A^M D_M-\frac{1}{d-2}(D_M A^M)+A^2 \Big]\, \psi\, .
\ee
In this formula~$D_M$ is the celebrated Thomas~$D$-operator which maps weight~$w$ tractors to weight~$w-1$ tractors ({\it i.e.} it respects the equivalence relation~\eqn{coneE}); in the ambient space it is given by the operator~$D_M:=\nabla_M (d+2\nabla_X-2)-X_M \Delta$. In  turn, this allows the action to be written as an Weyl invariant integral over the underlying conformal manifold~$M$
\begin{equation}\label{SgA}
S([g],A_M,\psi,\phi)={\rm Re}\int_M\, \overline{\phi\,} \Big[ \frac 1 w A^M D_M-\frac{1}{d-2}(D_M A^M)+A^2 \Big]\psi\, .
\end{equation}
The  integrand here depends on some metric~$g$ from the conformal class of metrics~$[g]$ on~$M$ determined by the FG metric~$g_{MN}$ and the integral is over the corresponding volume form. The product of the volume form and integrand above is Weyl invariant, so the action depends only on the conformal class of  the metric $[g]$, as indicated. Presently we will show that this action is in fact just a rewriting of the Einstein--Hilbert action. Before doing so,
we note that integrating out~$(\psi,\phi)$ in the path integral of this action gives the partition function
$$
Z=\int [Dg][DA_M] \left\{\det \Big[ \frac 1 w A^M D_M-\frac{1}{d-2}(D_M A^M)+A^2 \Big]\right\}^{-2}\, .
$$
Here to obtain a well-defined Gaussian and in turn a functional determinant, we  performed a Wick rotation on half the fields. 
This formula actually represents the partition function for the most {\it na\"ive}
 proposal for a model quantum gravity---an integration over metrics weighted by the exponential of the Einstein--Hilbert action---and thus should be compared with our proposal~\eqn{sumflatA0}.

Returning to our goal of obtaining the gravity action from~\eqn{psiQpsi},
we observe that the tractor-Maxwell field~$A_M$ in~\eqn{SgA} appears quadratically and algebraically so we can directly integrate it out.
In fact, the bottom slot of $A_M$ totally decouples and we have gauged the top slot to the constant $-w$. Thus we only need to algebraically determine the middle slot and find
$$
\bar A_\mu=-\, \frac{\overline \chi^T\! \sigma\,  \nabla_\mu \chi-\nabla_\mu\overline\chi^T  \sigma\,  \chi}{2\lambda}\, ,\qquad
\chi:=\begin{pmatrix}\psi\\ \phi\end{pmatrix}\, , \quad \lambda := \overline\chi^T \sigma \chi\, ,
$$
where $\sigma$ is the Pauli matrix $\sigma_x$ and the singlet $\lambda$ is Maxwell gauge invariant. 
Thus the on-shell covariant derivative becomes
$$
(\nabla_\mu+\bar A_\mu) \chi = {\bm \Pi} \, \nabla_\mu \chi + \frac{1}{2}\nabla_\mu \log \lambda\, \chi\, 
$$
where the projector
$$
{\bm \Pi}:={\mathbf 1}  \ -\ \frac{\chi \, \overline \chi^T \sigma}{\lambda}\, .
$$
Reinserting this in the action and calling $\lambda:=\varphi^2$ gives
$$
S([g],\chi)=\int_M\left[\nabla^\mu \overline\chi^T \sigma\,  {\bm \Pi}\,  \nabla_\mu\chi + \nabla^\mu \varphi \, \nabla_\mu \varphi +
\frac{d-2}{4(d-1)}\, R\,  \varphi^2
\right]\, .
$$
As indicated, this model only depends on the conformal class of the metric since it enjoys the gauge symmetry transformations
$$
g_{\mu\nu}\sim \Omega^2 g_{\mu\nu}\, ,\quad \chi \sim \Omega^{1-\frac d2} \chi \Rightarrow \varphi\sim \Omega^{1-\frac d2} \varphi \, .
$$
Note that the second two terms constitute the  action of a conformally improved scalar. One can  use that the projector
obeys ${\bm \Pi}\,  \chi = 0$ to verify conformal  invariance of the first term. The model therefore describes gravity coupled to a non-linear sigma model.
To see this, choose the gauge $\varphi = 1$ so that
$$
\overline \psi \phi + \overline \phi \psi = 1\, .
$$
This describes a hyperboloid in ${\mathbb R}^4$. 
Hence the action becomes, as promised,  a sum of the Einstein-Hilbert action plus  additional terms (with leading contribution a non-linear sigma model). Note, that without a Wick rotation of the scalar field measure, the Euclidean action has indefinite signs for its kinetic term. Given that we do not yet have full control over the moduli space of flat Chern--Simons connections, nor the integration measure for the underlying functional integral~\eqn{PartitionFunction} because the Hilbert space~${\cal H}$ is an indefinite relativistic one, it is premature to declare that the model has ghost excitations. We reserve a detailed study of this key issue to further work~\cite{BOW}.

\end{document}